\documentclass[aps,pra,epsfigure,twocolumn,superscriptaddress]{revtex4-2}%
\usepackage[colorlinks=true,linkcolor=blue,urlcolor=blue,citecolor=blue,pdfusetitle]{hyperref}
\usepackage[english]{babel}
\usepackage{amsmath}
\usepackage[caption = false]{subfig}
\usepackage{graphicx,epstopdf}
\usepackage{blindtext}
\usepackage[table,xcdraw,dvipsnames]{xcolor}
\usepackage{lipsum}
\usepackage{amsfonts}
\usepackage{bbm}
\usepackage{amssymb}
\usepackage{enumerate}
\usepackage{color}
\usepackage{latexsym}
\usepackage{physics}
\usepackage{times,txfonts}
\usepackage{soul}
\usepackage[normalem]{ulem}

\newcommand{\Ecal}{\mathcal{E}}

\newcommand{\Lcal}{\mathcal{L}}

\newcommand{\Ucal}{\mathcal{U}}

\newcommand{\1}{\mathbbm{1}}

\newcommand{\SubFig}[2]{\ref{#1}{\color{blue}#2}}

\newcommand{\mohammad}{\color{red}}
\usepackage{orcidlink}

\begin{document}

\title{Enhancing self-discharging process with disordered quantum batteries}

\author{Mohammad B. Arjmandi\!\!~\orcidlink{0000-0002-9222-6765}}
\email{m.arjmandi@sci.ui.ac.ir}
\affiliation{Department of Physics, University of Isfahan, P.O. Box 81746-7344, Isfahan, Iran.}
\affiliation{Quantum Optics Research Group, University of Isfahan, Isfahan, Iran}

\author{Hamidreza Mohammadi\!\!~\orcidlink{0000-0001-7046-3818}} 
\email{hr.mohammadi@sci.ui.ac.ir}
\affiliation{Department of Physics, University of Isfahan, P.O. Box 81746-7344, Isfahan, Iran.}
\affiliation{Quantum Optics Research Group, University of Isfahan, Isfahan, Iran}

\author{Alan C. Santos\!\!~\orcidlink{0000-0002-6989-7958}}
\email{ac\_santos@df.ufscar.br}
\affiliation{Departamento de Física, Universidade Federal de São Carlos, Rodovia Washington Luís, km 235 - SP-310, 13565-905 São Carlos, SP, Brazil}
\affiliation{Department of Physics, Stockholm University, AlbaNova University Center 106 91 Stockholm, Sweden}

\begin{abstract}
One of the most important devices emerging from quantum technology are quantum batteries. However, self-discharging, the process of charge wasting of quantum batteries due to decoherence phenomenon, limits their performance, measured by the concept of ergotropy and half-life time of the quantum battery. The effects of local field fluctuation, introduced by disorder term in Hamiltonian of the system, on the performance of the quantum batteries is investigated in this paper. The results reveal that the disorder term could compensate disruptive effects of the decoherence, i.e. self-discharging, and hence improve the performance of the quantum battery via "incoherent gain of ergotropy" procedure. Adjusting the strength of disorder parameter to a proper value and choosing a suitable initial state of quantum battery, the amount of free ergotropy, defined with respect to free Hamiltonian, could exceed the amount of initial stored ergotropy. In addition harnessing the degree of disorder parameter could help to enhance the half-life time of the quantum battery. This study opens perspective to further investigation of the performance of quantum batteries that explore disorder and many-body effects.
\end{abstract}

\maketitle

\section{Introduction}

Quantum mechanics holds the promise of a technological revolution, in which the successive miniaturization of electronics devices allow us to explore quantum effects such as quantum coherence and entanglement. There is a huge tendency to harness such effects in order to build quantum devices which possibly demonstrate an advantage over their classical counterparts. These devices range from quantum transistors~\cite{geppert2000quantum,Marchukov:16,joulain2016quantum,Loft:18,mcrae2019graphene,dePonte:19}, diodes~\cite{shukla2008nonlinear,shen2014quantum} and memories~\cite{lvovsky2009optical,hedges2010efficient,dennis2002topological} to the most popular field of nowadays, namely quantum computers~\cite{ladd2010quantum,divincenzo1999quantum}. So, one may ask whether it is also possible to construct an energy storage quantum system, i.e., a quantum battery (QB) as it was showed by Alicki and Fannes~\cite{Alicki:13}.

QBs are physical systems able to explore quantum effects and properties in order to provide an enhanced performance concerning their classical counterparts, in which an energy conversion occurs from chemical energy to electrical one, by means of the so-called reduction-oxidation reactions ~\cite{schmidt2018batteries,weber2011redox,pan2015redox}. In a QB, its ergotropy (maximum amount of extractable energy by unitary operations~\cite{Allahverdyan:04}) can be stored as entanglement~\cite{Santos:20c} and/or coherence ~\cite{Baris:20,Santos:21b}. But, it is also possible to explore these quantities in order to speed-up the charging process of such batteries~\cite{Binder:15,PRL2017Binder,PRL_Andolina,PRL2013Huber,araya2019geometrical}. However, it is important to mention that the role of entanglement is not positive in general~\cite{Kamin:20-2}, and it can lead to a destructive effect on the charging process, while quantum coherence seems to be a useful resource for QBs. More recently, the study of disordered QBs showed how disordered interactions between quantum cells can be used to promote the rate of charging power of QBs ~\cite{ghosh2020enhancement} by exploring many-body localization~\cite{Andolina:19-2}. Through this paper, the performance of the QB is measured by the concept of ergotropy and the notion of halfe-life time of QB.

Decohernce arising from the inevitable interaction of a QB (or any other quantum system) with its surrounding medium tends to bring the system to a non-active(passive) equilibrium state and hence deteriorates the performance of the QB~\cite{PRL2019Barra}. In this sense, the development of QBs which are robust against decoherence is an achievement of interest to build useful energy storing quantum devices. The charging process of QBs can be made robust in presence of decoherence when some degree of control and reservoir engineering is possible~\cite{Carrega:20}, or when the QB-environment interaction results into a non-Markovian dynamics~\cite{Kamin:20-1,Ghosh:21}. In addition, when the battery is charged it is necessary to make sure that the stored energy will not be lost to the environment, even when the battery is not connected to any consumption hub via a process called \textit{self-discharging}~\cite{Santos:21b}. The robustness of QBs against such a kind of process has been investigated experimentally in superconducting devices, suggesting that QBs exhibit a supercapacitor-like behavior during the self-discharging process~\cite{Santos:21b,Hu:21}.  

Given the role of disorder parameter, adjusted by applying random magnetic fields on each QB, to the charging performance of QBs, in this paper we wonder how the disorder can affect the self-discharging process~\cite{Santos:21b}, which is an open question so far. To this end, we consider a XX spin-1/2 chain with random magnetic fields on each cell of the QB. By considering two situations in which the initial amount of ergotropy is stored as quantum coherence or populations of classical states, our results show that in the presence of disorder the QBs are more robust against self-discharging when ergotropy is stored as coherence, leading then to a quantum advantage. The paper is organized as follows. In Sec.~\ref{SecII}, we present our model, method and figures of merit to investigate the self-discharging of QBs in the presence of disorder. Then, we show the results for two-cell batteries with different initial cell states. In Sec.~\ref{SecIII} we generalize our results to the case of $N$-cell batteries and represent the scaling trend of maximal ergotropy with the number of quantum cells and the degree of disorder. Finally, we conclude our results in Sec.~\ref{SecIV}.

 \section{ Quantum batteries in the presence of disorder} \label{SecII}

In this paper we  consider QBs as XX spin-$1/2$ chain in the presence of the disorder realized by random local magnetic fields $\vec{B}_{k}$ applying on $k$-th QB (along quantization axis $z$) described by the total Hamiltonian $H = H_{0}+H_{\mathrm{int}}$, where
\begin{align}
H_{0} = \sum_{k=1}^{N} \frac{\hbar\omega_{k}}{2}\sigma_{k}^{z} \label{Eq-FreeHamiltonian},
\end{align}
is the free Hamiltonian, used as reference Hamiltonian to compute ergotropy, and 
\begin{align}
H_{\mathrm{int}} = \sum_{k=1}^{N-1}\frac{\hbar J_{k,k+1}}{4} \left(\sigma_{k}^{x} \sigma_{k+1}^{x} + \sigma_{k}^{y} \sigma_{k+1}^{y}\right) \label{Eq-InteractionHamiltonian},
\end{align}
is the interaction Hamiltonian. Here, $\omega_{k}\!\propto\!|\vec{B}_{k}|$ is the Larmor frequency of the $k$-th spin, $J_{k,k+1}$ is the spin-spin nearest-neighbor interaction strength and $\sigma^{\alpha}$ $(\alpha\!=\!x,y,z)$ are the standard Pauli matrices. Thorough the paper we set $\hbar=1$ and assume the coupling strengths as $J_{k,k+1}=J$. The disordered property of the system is encoded into the magnetic fields acting on the $k$-th cell through the (dimensionless) strength of disorder $\delta$ as $B_{k}=\mathrm{random}(B_{0} - \delta B_{0},B_{0} + \delta B_{0})$, where $B_{0}$ is the reference field.

In order to investigate the performance of our system in keeping stored energy as ergotropy in a dissipative scenario, it is necessary to determine the dynamics of the system. Tracing out the environmental degrees of freedom over unitary evolution of the whole QB+environment in the Born-Markov approximation, leads to the Lindblad master equation for reduced state, $\rho$, of the QB:
\begin{align}
\frac{d\rho(t)}{dt}= -i [H,\rho(t)]+\frac{\Gamma}{2}\sum_{k}^{N}2 \sigma_{k}^{-} \rho(t) \sigma_{k}^{+} - \left\{\sigma_{k}^{+} \sigma_{k}^{-},\rho(t)\right\}, \label{Eq-MasterEq}
\end{align}
where the dissipative nature of the environment such as \textit{spontaneous decay} is encapsulated in the dissipation rate parameter, $\Gamma$ and ``$\{,\}$" denotes` the anti-commutator relation and $\sigma_{k}^{-}=(\sigma_{k}^{+})^{\dagger}=\1_{1}  \otimes...\otimes \1_{k-1} \otimes|g\rangle_{k}\langle e| \otimes \1_{k+1}...\otimes \1_{N}$ are raising and lowering operators with $|g\rangle$ and $|e\rangle$ being the ground and excited states of the system, respectively and $\1_{i}$ denotes $2\times2$ identity matrix. The physical meaning of the above master equation depends on the physical system we encode the quantum battery. If we have a spin quantum battery, the above equation describes the spin relaxation along quantization axis-$z$~\cite{Sarthour:Book}. In case the battery is given by atoms inside a cavity~\cite{Crescente:20}, the dynamics above describes the spontaneous atom decay to the cavity modes~\cite{Scully:Book}.  In a more general way, the above equation describes the dynamics of a system embedded in an environment modeled by a low-temperature bosonic bath, such that the temperature effects are negligible. In order to keep our discussion as the most generic as possible, we will not particularize our discussion to a specific quantum system. However, the validation regime of such an equation has been discussed in different physical systems~\cite{PhysRevA.55.2290,PhysRevA.59.1633,singh2020using,PhysRevA.94.033819,Fonseca_Romero_2012,Kockum:18,PhysRevLett.93.140404}. Discussion about the validity of such an equation can be found in Refs.~\cite{Petruccione:Book,Nielsen:Book}.

\subsection{Figures of merit}

The performance of a quantum battery in a charging or discharging process can be evaluated by means of various figures of merit. Naturally, the first potential candidate is the amount of stored energy. Consider a quantum battery with density matrix $\rho$ and Hamiltonian $H$. Then, the stored energy in this system with respect to its Hamiltonian is just the expectation value of $H$ given as $\Ucal(\rho,H)=\tr(H \rho)$.
In a general case, however, this energy can not be entirely extracted through external extracting fields (\textit{i.e.} applying unitary operations on $\rho$) ~\cite{Allahverdyan:04,Santos:20c}. In this regard, another important quantity is ergotropy, originated from quantum thermodynamics formulation of work extraction. The concept of ergotropy was initially introduced by Allahverdyan et al.~\cite{Allahverdyan:04} and determines the maximum amount of energy which can be extracted from a quantum system by unitary operations. A quantum system without the capability of unitarily energy extraction is called passive. Then, ergotropy is defined as $\mathcal E(\rho,H) = \Ucal(\rho,H)-\Ucal(\bar{\rho},H)$, in which $\bar{\rho}$ is the passive state associated with $\rho$. By definition, it is not possible to extract energy from a system that is in a passive state $\bar{\rho}$. Hence, the most successful energy extraction operations are those that transform the quantum state $\rho$ to $\bar{\rho}$. As it was shown in Ref. ~\cite{Allahverdyan:04}, let $\rho = \sum_{k=1}^{d} {p_{k} |p_{k} \rangle \langle p_{k}|}$ and $H = \sum_{k=1}^{d} {\epsilon_{k} |\epsilon_{k} \rangle \langle \epsilon_{k}|}$ to be the spectral decomposition of the battery state and Hamiltonian, with the conditions $p_{1}\geq p_{2}\geq . . . \geq p_{d}$ and $\epsilon_{1}\leq \epsilon_{2}\leq . . . \leq \epsilon_{d}$, where $d$ denotes the dimension of the Hilbert space, the ergotropy can be obtained as

\begin{align}
\mathcal E(\rho,H) = \sum\nolimits_{k,j}^{d} {p_{k} \epsilon_{j} (|\langle p_{k}|\epsilon_{j}\rangle|^{2} - \delta_{k,j})}, \label{Eq-Ergotropy}
\end{align}

where $\delta_{k,j}$ denotes the Kronecker delta function. It is known that if $\rho$ is a pure state, then the corresponding passive state is its ground state ~\cite{Santos:20c}. Throughout this paper, we focus on the free ergotropy which is calculated with respect to $H_{0}$ given by Eq.~\eqref{Eq-FreeHamiltonian}, namely, free ergotropy is defined as $\mathcal E(\rho,H_{0})$. In this sense, the free ergotropy quantifies the amount of energy that can be (individually) extracted from each cell, since any contribution (positive or negative) of the interactions does not affect $\mathcal E(\rho,H_{0})$. 

It is worth mentioning that, due to the random nature of our model, all calculations should be repeated and then the physical quantities shall be averaged over the number of realizations $N_{r}$. In this sense, we define the average ergotropy as
\begin{align}
\overline{\mathcal E}(t)=\frac{1}{N_{r}} \sum\nolimits_{n=1}^{N_{r}}{\mathcal E(\rho^{n}(t),H_{0}^{n})},
\end{align}
where $H_{0}^{n}$ and $\rho^{n}(t)$ are, respectively, the free Hamiltonian and time-depended state of the battery obtained for $n$-th realization. In addition, in order to avoid any misinterpretation in our analysis, since the random aspect of local magnetic $\vec{B}_{k}$ will change the maximum amount of ergotropy in each random realization, we define our figure of merit as
\begin{align}
\varepsilon(t)= \frac{1}{N_{r}} \sum\nolimits_{n=1}^{N_{r}}\frac{\mathcal{E}(\rho^{n}(t),H_{0}^{n})}{\mathcal{E}(\rho^{n}(0),H_{0}^{n})},\label{Eq-NormErgo}
\end{align}
which quantifies how the total ergotropy evolves concerning the initially stored ergotropy in each realization $\mathcal{E}(\rho^{n}(0),H_{0}^{n})$.

\subsection{Quantum advantage against self-discharging process}

\begin{figure}[t!]
\includegraphics[width=\linewidth]{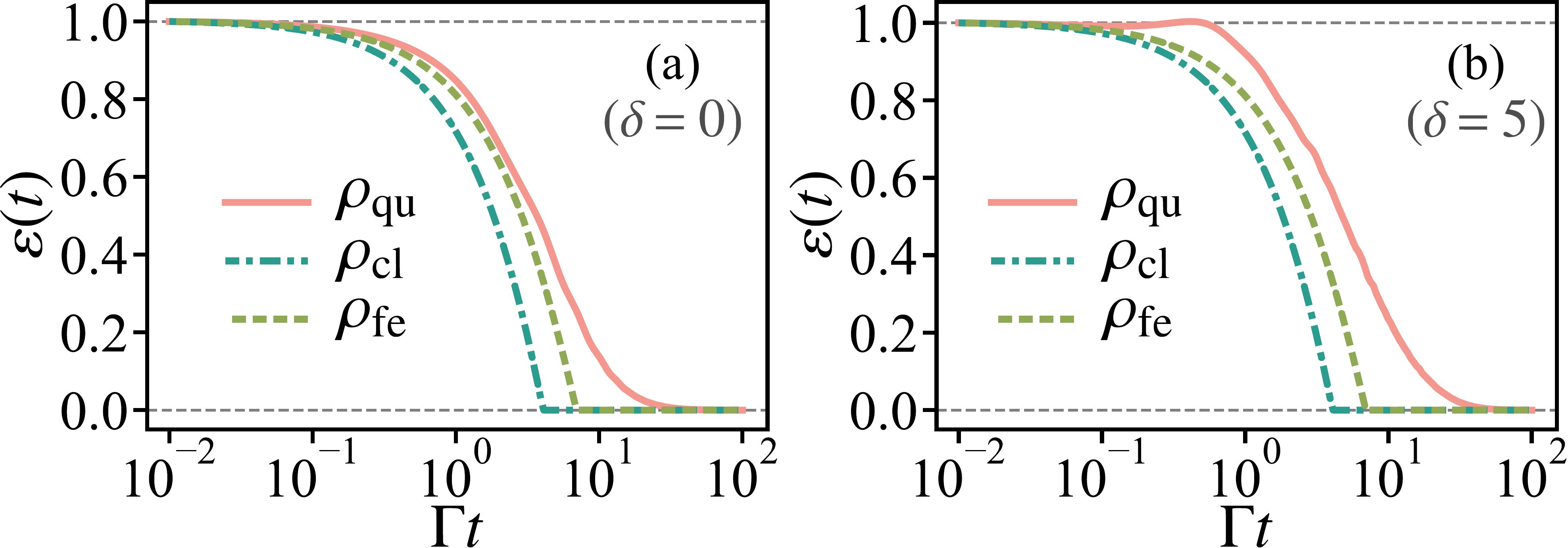}
\caption{The dynamics of averaged ergotropy of the two-cell QBs ($N=2$) with different initial state (a) in the absence of disorder and (b) in the presence of disorder. The results are averaged over $10^{2}$ realizations and $J_{k,k+1}=J=10 \Gamma$.}
\label{Fig-Graph1}
\end{figure}

\begin{figure}[t!]
\includegraphics[width=\linewidth]{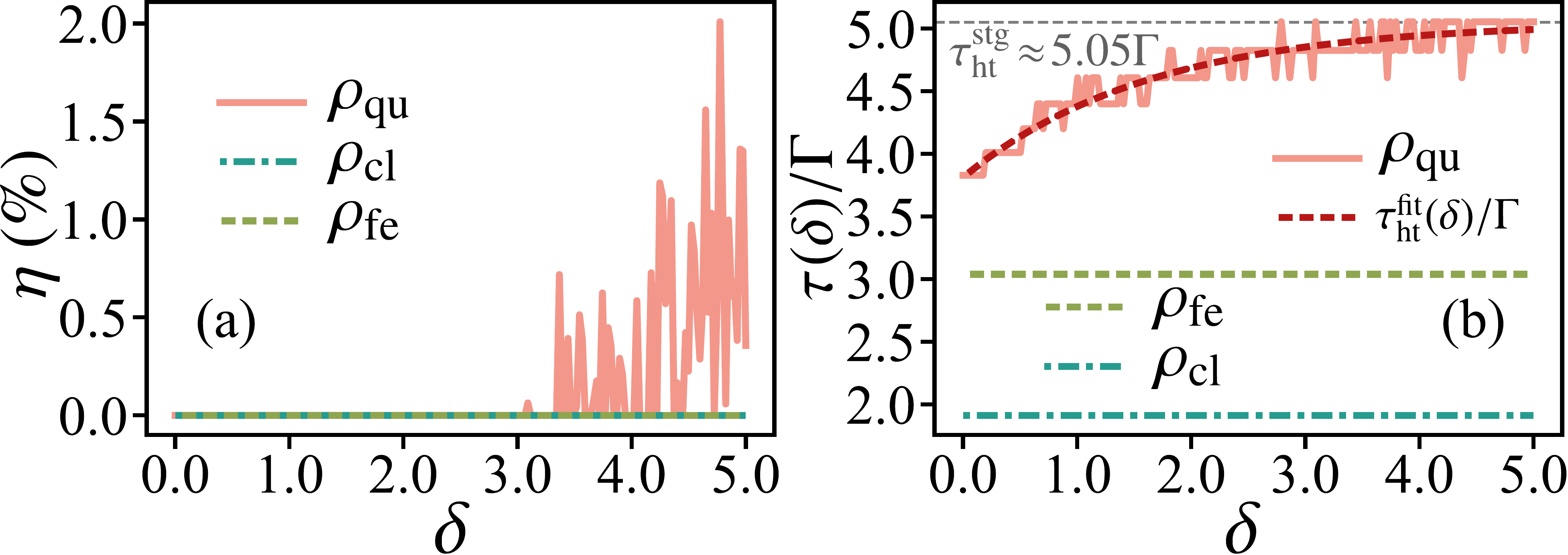}
\caption{(a) The percentage of spontaneous deposited charge and (b) half-life time of the three batteries versus the degree of disorder. The other parameters are the same as Fig.~\ref{Fig-Graph1}.}
\label{Fig-Graph2}
\end{figure}

In order to analyze the self-discharging process of a two-cell QB in the presence of disorder and how the quantum advantage arises in this process, we compare three different initial states used to store an initial amount ergotropy. First, we assume the fully charged scenario, where the initial state  of QB is $\rho_{\mathrm{fe}}\!=\!\ket{ee}\bra{ee}$. We also consider the battery in the state $\rho_{\mathrm{qu}}\!=\!\ket{++}\bra{++}$, where $|+\rangle\!=\!(|g\rangle+|e\rangle)/\sqrt{2}$, such that ergotropy is stored as local quantum coherence in the basis $\{\ket{g},\ket{e}\}$~\cite{Baris:20}. This state posses local coherence with respect to the basis $\{\ket{g},\ket{e}\}$, then we call this state as \textit{"coherent initial state"} through this paper. Finally, we also consider a classical mixture of the states $\ket{g}$ and $\ket{e}$ as $\rho_{\mathrm{cl}}\!=(\!\alpha|e\rangle \langle e| + \beta |g\rangle \langle g|)^{\otimes 2}$, where we set $\alpha\!=\!3/4$ to make sure that the quantum version of the battery has the same amount of stored ergotropy as its classical counterpart~\cite{Santos:21b}. It is worth to mention that our model does not present any kind of \textit{quantum supremacy}. As shown in Ref.~\cite{Santos:21b}, any advantage against self-discharging of quantum states can be lost if you consider other kinds of decoherence. The results of numerical calculation are depicted in Figs. ~\ref{Fig-Graph1}-~\ref{Ap-Appen-RatioN=7}.

In the Fig.~\SubFig{Fig-Graph1}{a} we show the average normalized ergotropy (Eq.~\eqref{Eq-NormErgo}) in the absence of disorder ($\delta\!=\!0$) and for the non-zero degree of disorder ($\delta\!=\!5$) in the case of two-cell ($N\!=\!2$) quantum battery for each initial state aforementioned. In both cases, the coherent initial state represents the best performance in terms of self-discharging, compared to the classical and full excited states. This advantage of the coherent initial state over the two other ones is even increased in the presence of disorder. Actually, the classical and full excited states are not sensitive to disorder.

As one can see from Fig.~\SubFig{Fig-Graph1}{b}, when the available work is stored as coherence in the system presents a robustness against the spontaneous loss of ergotropy. Moreover, it is worth to highlight the behavior of the ergotropy that, after an small negative variation, presents an slight positive variation in the presence of disorder before its monotonic decreasing (expected) behavior. Since the ergotropy variation in our system is due to the spontaneous decay considered in our system, and no external field is acting on the system, we call this behavior \textit{incoherent gain of ergotropy} of the quantum battery. In addition, it is possible to show that this behavior is associated to the degree of disorder and it can be explained as follows. Let us consider the free ergotropy written in the following form $\Ecal(\rho,H_{0})\!=\!\Ucal(\rho,H_{0}) - \Ucal_{0}(\rho,H_{0})$, where $\Ucal_{0}(\rho,H_{0})\!=\!\sum_{k}p_{k}(t)\epsilon_{k}$ and the \textit{free} internal energy $\Ucal(\rho,H_{0})\!=\!\tr(\rho H_{0})$ is computed concerning the free Hamiltonian $H_{0}$~\cite{Luiz:21}. Then, by taking the time variation of $\Ecal(\rho,H_{0})$ in an infinitesimal time interval $dt\!\geq\!0$, we get
\begin{align}
d\Ecal(\rho,H_{0}) = \left[\frac{d\Ucal(\rho,H_{0})}{dt} - \frac{d\Ucal_{0}(\rho,H_{0})}{dt}\right]dt . \label{Eq-dE}
\end{align}
From above equation we can identify different situations where we get $d\Ecal(\rho,H_{0})>0$, independent of the process we are dealing with, which leads to the increasing behavior of the ergotropy. All these cases can be unified in a single condition given by $\dot{\Ucal}(\rho,H_{0})\!>\!\dot{\Ucal}_{0}(\rho,H_{0})$. In order to give an physical meaning to this condition, let us write temporal derivative of the internal energy as
\begin{align}
\dot{\Ucal}(\rho,H_{0}) = \tr(\dot{\rho} H_{0})=-i \tr(\rho(t)[H_{0},H_{\mathrm{int}}]) + \tr(\Lcal[\rho(t)]H_0) , \label{Eq-DUH0}
\end{align}
where $\Lcal[\rho(t)]$ is the dissipative part of the Eq.~\eqref{Eq-MasterEq}. Then the system evolves and the above quantity is such that we get $\dot{\Ucal}(\rho,H_{0})\!>\!\dot{\Ucal}_{0}(\rho,H_{0})$, then we have a gain of ergotropy. Therefore, any property of the system which positively contributes to $\dot{\Ucal}(\rho,H_{0})$, can be helpful to charge the battery. In our system of interest, it is possible to conclude that in the absence of disorder we have $[H_{0},H_{\mathrm{int}}]\!=\!0$, which means  only the dissipative term contributes to the inequality $\dot{\Ucal}(\rho,H_{0})\!>\!\dot{\Ucal}_{0}(\rho,H_{0})$. Otherwise, when we turn ON the disorder in our system, the first term of the Eq.~\eqref{Eq-DUH0} is not vanishing and it may positively (or negatively) contribute to achieve a regime in which $\dot{\Ucal}(\rho,H_{0})\!>\!\dot{\Ucal}_{0}(\rho,H_{0})$. The results of the Fig.~\ref{Fig-Graph1} suggest that, in average over many realizations, the disorder is positively contributing to get the inequality $\dot{\Ucal}(\rho,H_{0})\!>\!\dot{\Ucal}_{0}(\rho,H_{0})$. As shown in Fig.~\SubFig{Fig-Graph1}{b}, it is possible to identify situations in which the instantaneous (average) ergotropy becomes bigger than the initial stored ergotropy, which means that the battery is charged due to the interaction with its surrounding. As we shall see, this behavior becomes even more significant when we have many cells QBs, where the positive contribution of the first term of the Eq.~\eqref{Eq-DUH0} to achieve the inequality $\dot{\Ucal}(\rho,H_{0})\!>\!\dot{\Ucal}_{0}(\rho,H_{0})$ is stronger for $N=7$ than the case $N=2$.
In addition, in Fig.~\SubFig{Fig-Graph2}{a} we show the amount of spontaneously charged ergotropy (as a percentage of the initial stored ergotropy) as a function of the disorder, which is mathematically defined as (in percentage terms)
\begin{align}
\eta = \left[\frac{1}{\varepsilon(0)}\left(\max_{t>0} \varepsilon(t)\right) - 1 \right] \times 100. \label{Eq-eta}
\end{align} 

While classical and fully-excited states does not present a time interval where the incoherent gain of ergotropy appears, the quantum version can be benefited in a disordered system. It is worth to highlight that this result has no any contradiction with energy conservation, since the ergotropy increasing process is associated to the free Hamiltonian $H_{0}$. In fact, for the case where the reference Hamiltonian as given by $H\!=\!H_{0}+H_{int}$, we expect that the available work as ergotropy follows a monotonically decreasing behavior in time during the self-discharging process.

As a second figure of merit, we explore the disorder effects in the half-life time of the quantum battery. By definition, we refer to half-life time of a battery as the longest time interval $t:[0,\tau_{\mathrm{ht}}]$ for which the battery retains at least $50$ percent of its initial charge value.
Our definition is based on fact that a quantum battery does not decay following an Ohmic behavior~\cite{Hu:21,Santos:21b}, then the ergotropy is not well described by an exponential decay. The Fig.~\SubFig{Fig-Graph2}{b} shows the half-life time of three quantum batteries against disorder. As result, we can see that quantity half-life time scales with the degree of disorder for quantum battery in the coherent initial state, and a fitted curve is provided to clarify the scaling trend. From the fitting function, defined here as $\tau_{\mathrm{ht}}^{\mathrm{fit}}(\delta)/\Gamma\!=\!\alpha + \beta e^{ -\gamma \delta }$, we get the set of fitting parameters as $\alpha\!\approx\!5.05$, $\beta\!\approx\!-1.24$ and $\gamma\!\approx\!0.61$. In this form, we can determine the half-life time for a strongly disordered system ($\delta\!\gg\!1$) as $\tau_{\mathrm{ht}}^{\mathrm{stg}}\!=\!\alpha\Gamma$. It shows that the disorder can broaden the time scale for which $\Ecal(t)\!\geq\!\Ecal(0)/2$, with a threshold bounded (in average) by $\tau_{\mathrm{ht}}^{\mathrm{stg}}\!\approx\!5.05\Gamma$. In conclusion, the classical and fully-excited batteries are deprived of such advantage, although the performance of fully-excited battery is better than the classical one.

\section{N-cell quantum batteries in the presence of disorder} \label{SecIII}

\begin{figure}[t!]
\includegraphics[width=\linewidth]{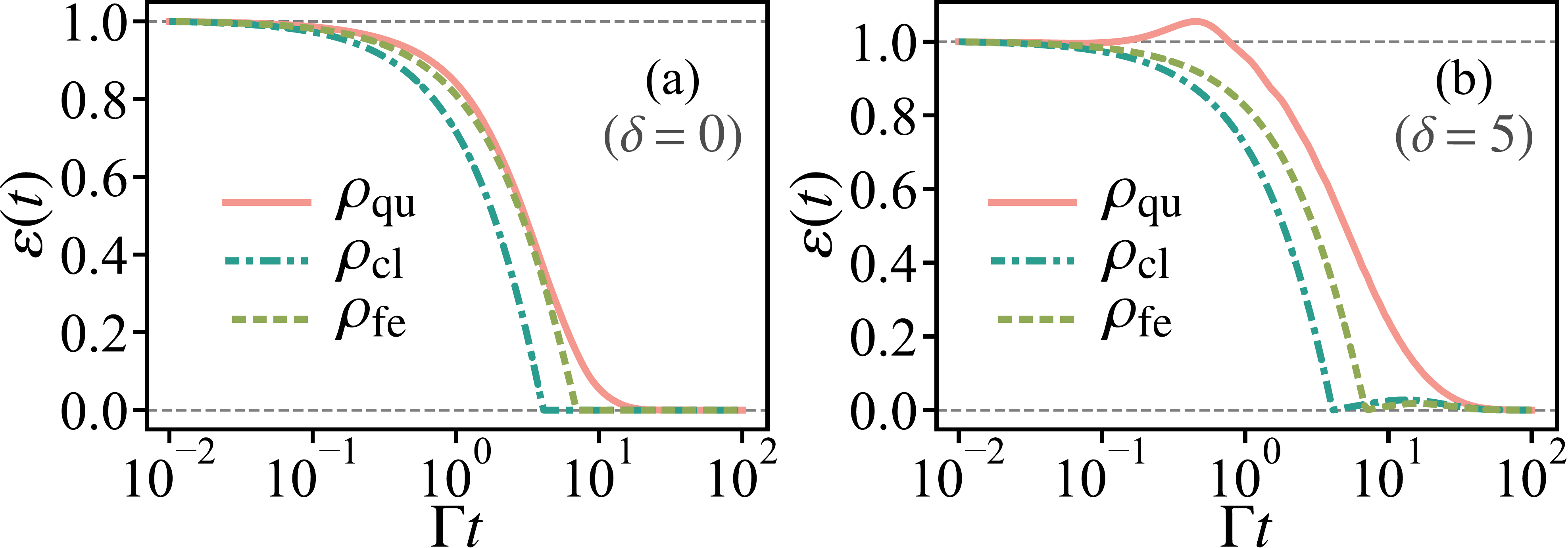}
\caption{The dynamics of averaged ergotropy of the multi-cell QBs ($N=7$) with different initial state  (a) in the absence of disorder and (b) in the presence of disorder. The results are averaged over $10^{2}$ realizations and $J_{k,k+1}\!=\!J\!=\!10 \Gamma$.}
\label{Fig-Graph3}
\end{figure}

In this section, we extend our previous results to the case of $N$-cell quantum battery, in order to see whether the disorder-based advantage also exists in larger quantum batteries. After such an analysis, one can verify the way by which this advantage scales when the system is enlarged enough. By changing the battery size from $N\!=\!2$ to $N\!=\!7$, we can see an enhancement concerning the spontaneous incoherent charging. This result is shown in Fig.~\ref{Fig-Graph3}, where we present a comparison between our three initial states for ordered ($\delta=0$) and disordered ($\delta=5$) cases (similar to the Fig.~\ref{Fig-Graph1} with $N\!=\!7$). By focusing on the case without disorder, although the coherent initial state again appears to have the best performance among all the initial states, its performance against spontaneous decay is almost similar to its two-cell counterpart. It can be seen by a direct comparison between the Figs.~\SubFig{Fig-Graph3}{a} and ~\SubFig{Fig-Graph1}{a}. In fact, in spite of identical spontaneous decay rate, the number of intracell interaction increases when we have $N\!=\!7$ compared to $N\!=\!2$, so they experience a slightly different dynamics from each other. By intracell interaction we  refer to the number of interaction between the cells of the quantum battery. 

\begin{figure}[t!]
\includegraphics[width=\linewidth]{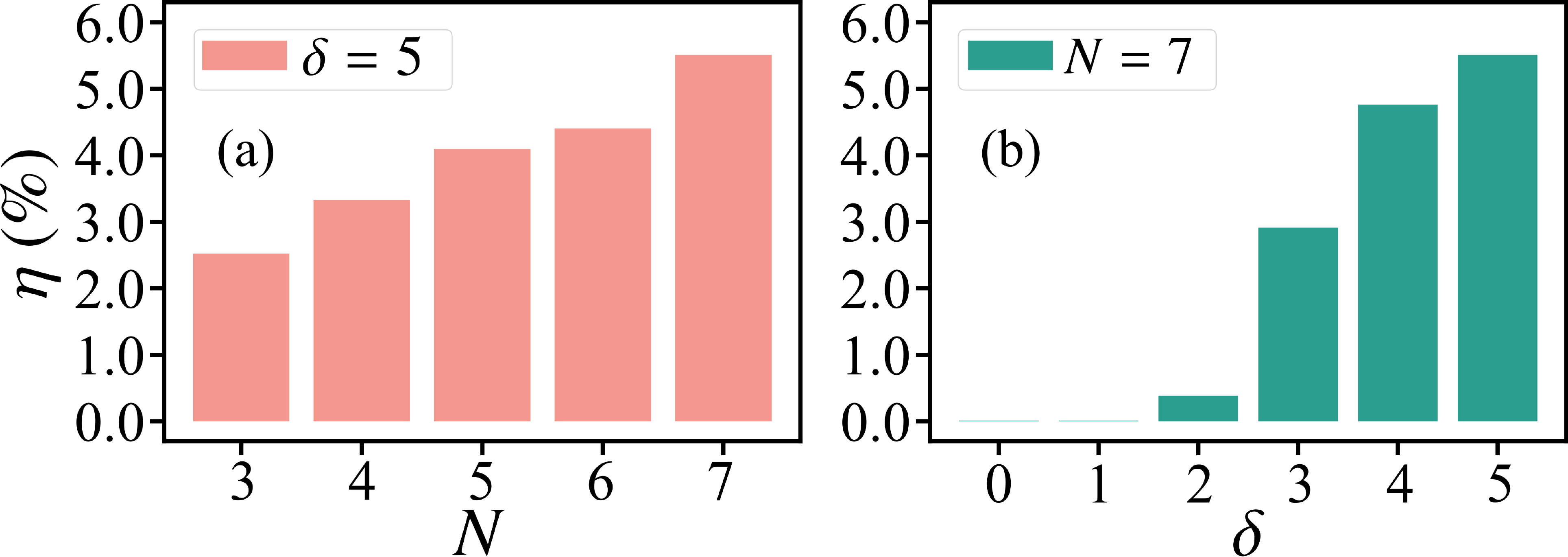}
\caption{The percentage of maximal spontaneous deposited charge of the coherent battery for (a) fixed degree of disorder ($\delta=5$) and different numbers of quantum cells and (b) fixed number of quantum cells ($N\!=\!7$) and various degrees of disorder. The number of realizations for $N\!=\!3,4,5,6,7$ are 5000, 3000, 1000, 500, 100, respectively and $J_{k,k+1}\!=\!J\!=\!10 \Gamma$.}
\label{Fig-Graph4}
\end{figure}

On the other hand, by implementing an appropriate amount of disorder, the behavior becomes different. As shown by Fig.~\SubFig{Fig-Graph3}{b}, we can make the battery more robust against the self-discharging process by implementing a certain amount of disorder in the system. This advantage is even more evident if one considers the amount of spontaneous ergotropy charging of the initially coherent battery (in comparison with its $N=2$ counterpart). This suggests that both the disorder parameter and number of intracell interactions play an important role in the performance of the system in keeping ergotropy stored. Then, in order to present a more adequate approach to this claim, the Fig.~\ref{Fig-Graph4} presents two graphs that show the relative gain of ergotropy concerning the initial amount of ergotropy, as given by Eq.~\eqref{Eq-eta}.

\begin{figure}[t!]
\includegraphics[width=\linewidth]{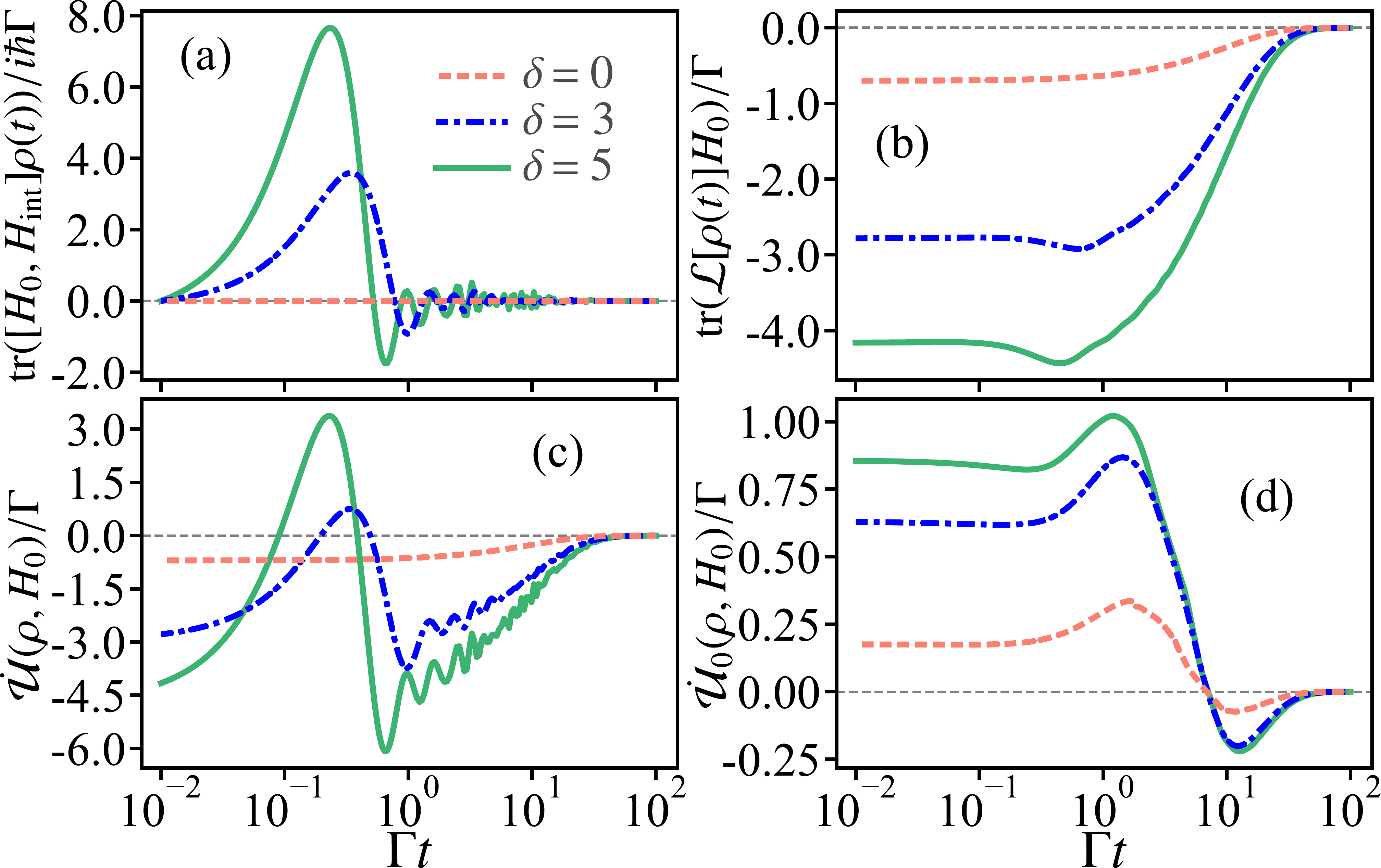}
\caption{ (a) The first and (b) second term of Eq.~\eqref{Eq-DUH0}, (c) time derivative of the internal energy and (d) $\dot{\Ucal}_{0}(\rho,H_{0})$ for $N=7$ and three degrees of disorder. The other parameters are the same as Fig.~\ref{Fig-Graph1}.}
\label{New-Fig}
\end{figure}

Now, in order to explain the incoherent generation of ergotropy, we show the contribution of each quantity defined in Eqs.~\eqref{Eq-dE} and~\eqref{Eq-DUH0}. As observed from Fig.~\SubFig{New-Fig}{a}, the first term of Eq.~\eqref{Eq-DUH0} grows considerably when disorder exists in the system, while in the absence of disorder it is vanished over the time interval. On the other hand, the dissipative part of the dynamics presents a negative contribution, which is expected since such term describes the loss of energy from the system to the environment. As result of the behavior of these two terms, a positive value of the quantity $\dot{\Ucal}(\rho,H_{0})$ is achieved (Fig.~\SubFig{New-Fig}{c}), which subsequently leads to the gain of ergotropy since it satisfies $\dot{\Ucal}(\rho,H_{0})>\dot{\Ucal}_{0}(\rho,H_{0})$ (Fig.~\SubFig{New-Fig}{d}). In conclusion, under disorder effects the first term of the Eq.~\eqref{Eq-DUH0} provides a strong contribution to the gain of ergotropy.

For a fixed degree of disorder, by increasing the number of intracell interactions ($N$) the relative gain is enhanced {\mohammad (Fig.~\SubFig{Fig-Graph4}{a})}. In fact, if the spontaneous ergotropy generation is associated to the individual emission of each partition of the system, it is reasonable to envision such a increasing behavior of $\eta$ as $N$ increases. In addition, in order to shed more light into the role of disorder regarding this incoherent ergotropy generation, in {\mohammad Fig.~\SubFig{Fig-Graph4}{b}} we present a scenario in which the number of quantum cells is fixed ($N=7$)  and the degree of disorder is varied. Here, without disorder ($\delta=0$) or with a slight amount of disorder ($\delta=1$), the maximal ergotropy is just equal to initial value before the energy leakage into environment, leading to $\eta=0$. However, raising the degree of disorder ($\delta>1$) results in a growth of maximum amount of ergotropy (approximately) scaling with disorder. As a general conclusion of the Fig.~\ref{Fig-Graph4}, it is possible to achieve an incoherent production of ergotropy of, at least, $5\%$ compared to the initial amount of stored ergotropy $\varepsilon(0)$ in disordered quantum batteries.

This discussion allows us, \textit{in principle}, to conjecture a strong dependence of the self-discharging performance  on the disorder and number of intra-cell interactions. We highlight here the term `in principle', since the many-body properties of quantum systems can lead us through non-trivial dynamics when the system is large enough. For example, it is perfectly possible to get an asymptotic behavior of $\eta$ for some larger values of $N$. Again, we stress that such a generation of ergotropy does not violate any physics law. For example, from Eq.~\eqref{Eq-dE} one can see that ergotropy can be generated from a nonzero amount of internal energy, and it can be converted into ergotropy during our dynamics.

\begin{figure}[t!]
\includegraphics[width=\linewidth]{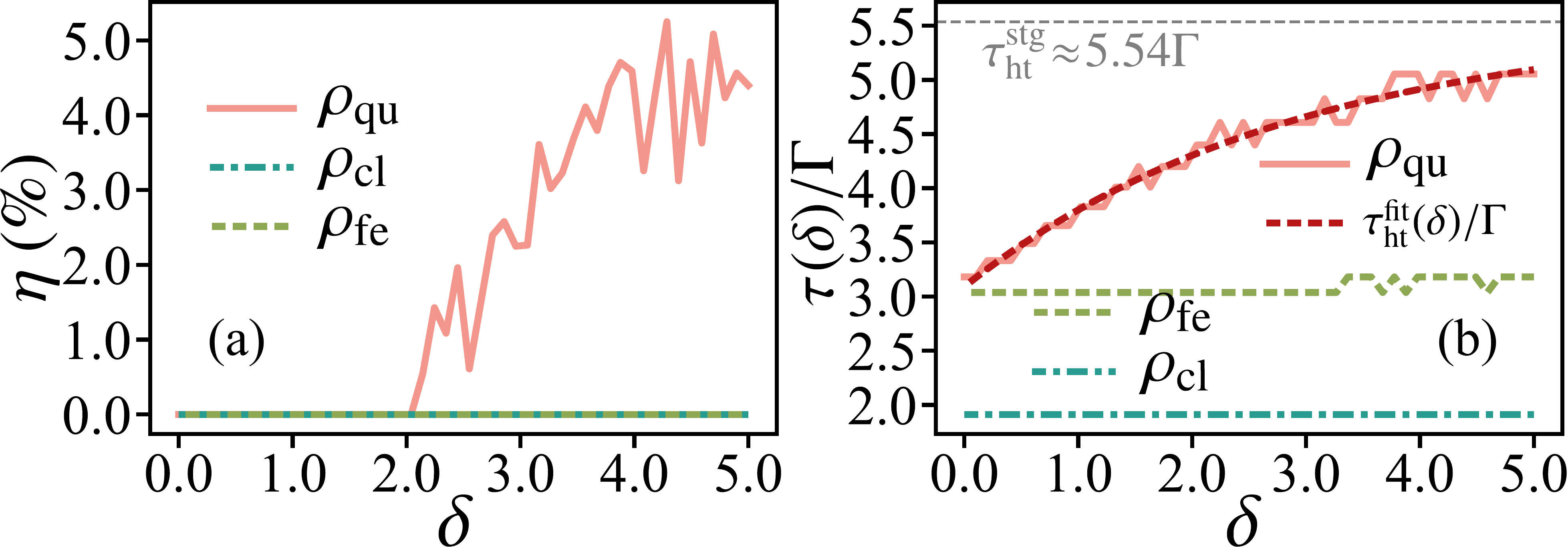}
\caption{(a) The percentage of spontaneous deposited charge and (b) half-life time of multi-cell batteries ($N\!=\!7$) versus the degree of disorder over $50$ realizations. Here also $J_{k,k+1}\!=\!J\!=\!10 \Gamma$.}
\label{Fig-Graph5}
\end{figure}

Finally, let us consider the effect of disorder on the maximum ergotropy and half-life time of $N$-cell batteries with different initial states. Fig.~\ref{Fig-Graph5} shows similar quantities as Fig.~\ref{Fig-Graph2}, but for $N=7$. From Fig.~\SubFig{Fig-Graph5}{a} it can be understood that smaller values of disorder ($\delta>2$) are able to promote the maximal ergotropy, compared to the two-cell quantum battery. It means that for a multi-cell quantum battery with initial coherence, a moderate degree of disorder can lead to an increment in deposited ergotropy though being exposed to a dissipative environment. But this is not the case for the  quantum batteries prepared in classical an/or fully-excited initial states, as there is no dependence on disorder for these states. The half-life of the battery is shown in Fig.~\SubFig{Fig-Graph5}{b}, where we again take the aforementioned fitting function to estimate the behavior of $\tau_{\mathrm{ht}}^{\mathrm{fit}}(\delta)$. For the $N$-cell QB one finds the set of fitting parameters as $\alpha\!\approx\!5.54$, $\beta\!\approx\!-2.44$ and $\gamma\!\approx\!0.34$, such that we get the higher disorder threshold bounded by $\tau_{\mathrm{ht}}^{\mathrm{stg}}\!\approx\!5.54\Gamma$ than two-cell case. More precisely, it represents an enhancement of $9.6\%$ with respect to the two-cell case.

It is important to mention that in our model we assume that the ergotropy of the battery is computed with respect to $H_0$ as the reference Hamiltonian~\cite{Gherardini:19,Andolina:19-2,PRB2019Batteries,Andolina:19,PRL2017Binder,Rossini:20,PRL2019Barra,Alicki:13,Crescente:20}, such that intracell interaction needs to be turned OFF before the ergotropy extraction process. The energy cost to disable the intracell interaction depends on the physical system used to store ergotropy. For example, when we have a spin-based QB such a cost has been considered in Refs.~\cite{Hovhannisyan:20,Cruz:22} in different scenarios, where the energy required to do that is bigger than the available amount of ergotropy, since we need strong local external fields to make the interaction negligible. However, when this interaction between the quantum cells is generated by light-mediated interactions the energy cost to turn OFF the interaction does not need to be taken into account. For example, in a Dicke quantum battery~\cite{Crescente:20,Dou:22}, where $N$ two-level systems interact with each other mediated by a single cavity mode, the intracell interaction is suppressed by eliminating the field inside the cavity~\cite{Mlynek:14,Davis:19}. In the best scenario, the energy cost of our battery can be considered the same as the Dicke QB model~\cite{Crescente:20,Dou:22}.

\section{Conclusions} \label{SecIV}

In this paper we study a quantum battery (QB) that can be mapped into any physical system constituted of two-level systems modeled by a XX Heisenberg model under spontaneous emission, in which disorder can be applied using random external local fields. In particular, we considered the role of disorder effects on the performance against self-discharging process of QBs. In order to explore quantum advantages, the ergotropy has been initially stored as quantum coherence and classical populations (from a classical mixed state and a fully-excited one).  Although the classical and fully-excited initial states are insensitive to disorder effects, when ergotropy is initially stored as coherence, a considerable disorder-induced enhancement arises in the performance of the  QB. We identify two main effects, where (i) the disorder can make the coherent quantum battery more robust against self-discharging, and (ii) a gain of ergotropy of the QB even under decoherence. It has been shown that degree of disorder and the number of quantum cells both contribute in this enhancement, where our results suggest that disorder can extend the half-life time of the battery with initial coherence. It implies that with an appropriate amount of disorder, one can significantly prolong the self-discharging process of the QB.

We would like to stress that we cannot give a general claim of this result due to the limited value of $N$ we consider here. In case our claim can be verified, it is possible to say that the larger the battery is, the greater the maximum ergotropy will be. As an immediate question that arises from our discussion, we wonder what would happen if we explore the self-discharging using different kinds of many-body effects. Given the recent experimental observation of many-body localization in an quantum annealer~\cite{Jaime:21}, our results motivate the possibility to propose new kind of quantum batteries.

\begin{acknowledgments}
A.C.S. acknowledges the financial support of the São Paulo Research Foundation (FAPESP)  through Grant Nos. 2019/22685-1 and 2021/10224-0.
\end{acknowledgments}

\begin{figure}[t!]
\includegraphics[width=\linewidth]{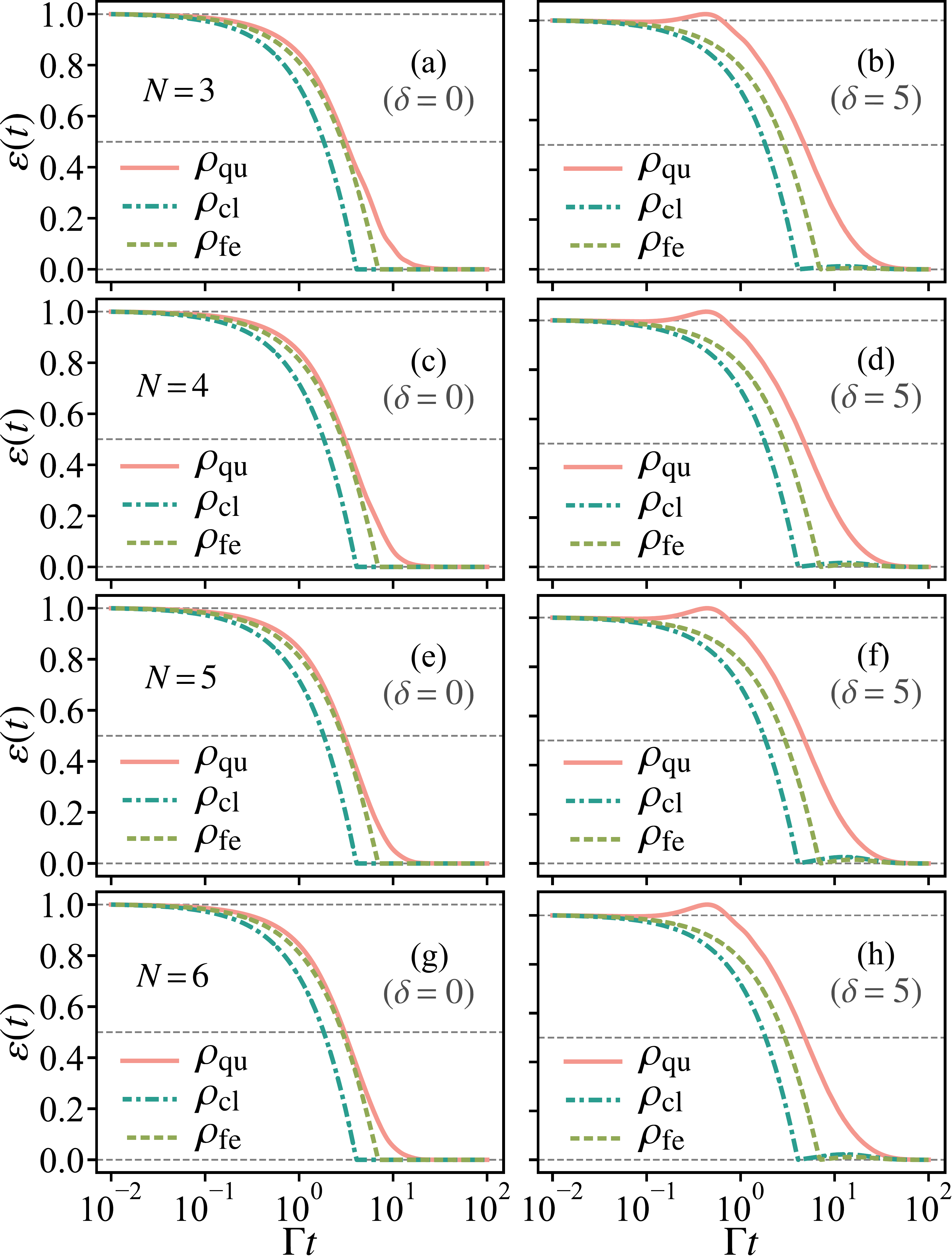}
\caption{Dynamics of ergotropy  of QBs in the absence (left) and presence  (right) of disorder  with various number of quantum cells. The results for $N=3,4,5,6$ are averaged over 5000, 3000, 1000 and 500 realizations, respectively. Here we consider $J_{k,k+1}\!=\!J\!=\!10 \Gamma$.}
\label{Ap-App-DifferentNs}
\end{figure}

\appendix


\section{Performance of the QB with respect to $N$}
Here we provide complementary results corresponding to performance of the QB with respect to $N$. Fig. ~\ref{Ap-App-DifferentNs} shows the dynamics of the averaged ergotropy for different number of quantum cells ($N\!=\!3,4,5,6$). As previously claimed, when the battery contains initial coherence its ergotropy tends to be more robust against self-discharging, storing a long-lived amount of energy in comparison to classical and fully-excited cases. The performance of the coherent battery is enhanced in the presence of disorder. Moreover, the temporary growth of ergotropy is evident again in disordered coherent quantum battery for all number of quantum cells. This result suggests that ergotropy in this kind of battery is able to be enhanced no matter how big the system is. Of course, as shown by FIG. ~\ref{Fig-Graph4} the amount of such enhancement scales with the size of quantum battery (at least up to $7$ cells).

\section{Stored energy \textit{versus} extractable energy (stored ergotropy)}

In this appendix we present a complementary discussion to the results shown in Figs.~\ref{Fig-Graph1} and~\ref{Fig-Graph3}. The stored energy is defined here as the internal energy of the system concerning the reference Hamiltonian, i.e., $\Ucal(\rho,H_{0})\!=\!\tr(\rho H_{0})$, while the extractable energy is given by the stored ergotropy defined in Eq.~\eqref{Eq-Ergotropy}. In order to quantify which portion of the stored energy corresponds to the extractable one, we define the ratio
\begin{align}
\Xi(t) = \frac{1}{N_{r}} \sum\nolimits_{n=1}^{N_{r}}\frac{\mathcal{E}(\rho^{n}(t),H_{0}^{n})}{\Ucal(\rho^{n}(t),H_{0}^{n})} , \label{Ap-Eq-NormErgo}
\end{align}
where the quantities $\mathcal{E}(\rho^{n}(t),H_{0}^{n})$ and $\Ucal(\rho^{n}(t),H_{0}^{n})$ are evaluated for each realization and we take the average of the ratio $\mathcal{E}(\rho^{n}(t),H_{0}^{n})/\Ucal(\rho^{n}(t),H_{0}^{n})$ over $N_{r}$ realizations. In order to get a comparative with the average stored energy, we define the normalized internal energy
\begin{align}
u(t) = \frac{1}{N_{r}} \sum\nolimits_{n=1}^{N_{r}}\frac{\Ucal(\rho^{n}(t),H_{0}^{n})}{\Ucal(\rho^{n}(0),H_{0}^{n})}.  \label{Ap-Eq-NormIntEner} 
\end{align}

\begin{figure}[t!]
\includegraphics[width=\linewidth]{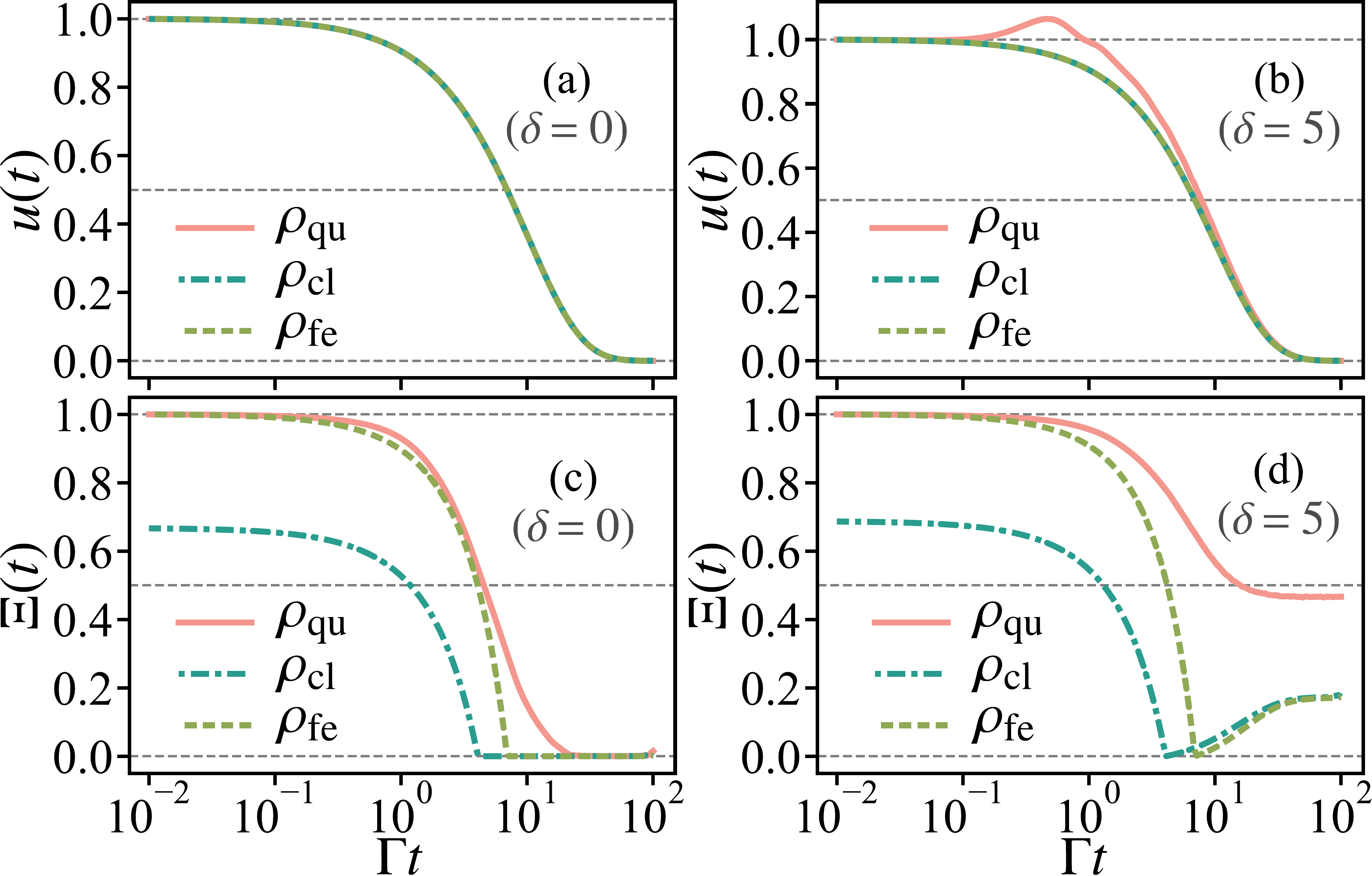}
\caption{The time evolution of internal energy and the ratio defined by ~\eqref{Ap-Eq-NormErgo} in the absence ((a) and (c)) and presence ((b) and (d)) of disorder with $N\!=\!7$. The results are averaged over 100 realizations . Here we consider $J_{k,k+1}\!=\!J\!=\!10 \Gamma$.}
\label{Ap-Appen-RatioN=7}
\end{figure}

Fig.~\ref{Ap-Appen-RatioN=7} shows the internal energy and the ratio of extractable energy for both ordered and disordered cases. Without disorder ($\delta=0$) no difference between the batteries with different initial state appears regarding the internal energy (Fig.~\SubFig{Ap-Appen-RatioN=7}{a}). In other words, in this case the stored energy in all of these considered batteries is dissipated by environment with equal rates. However, as seen from Fig.~\SubFig{Ap-Appen-RatioN=7}{c} the quantum battery with initial coherence possesses a larger ratio of extractable energy during the time interval. Based on previous results one could expect this behavior since the coherent battery turns out with better performance in terms of ergotropy even if no disorder exists. On the other hands, exposing some degree of disorder results in a temporal increase in internal energy of the coherent battery (Fig.~\SubFig{Ap-Appen-RatioN=7}{b}), just similar to its ergotropy enhancement. However, no considerable advantage against self-discharging of the stored energy is observed as the internal energy of the three batteries vanishes at the same point. But this is not the case for the ratio of extractable energy. As shown by Fig.~\SubFig{Ap-Appen-RatioN=7}{d}, initial coherence along with disorder can provide a larger ratio of extractable energy defined by ~\eqref{Ap-Eq-NormErgo} compared to the classical and fully-excited batteries.


%

\end{document}